\documentclass[
superscriptaddress,
 aps,
twocolumn,
pra,
nofootinbib,
]{revtex4-2}

\usepackage{tikz} 
\usepackage{amsmath,amsfonts,amsthm, mathtools}
\usepackage{float}
\usepackage{dsfont}
\usepackage{csquotes}
\usepackage{amssymb}
\usepackage{verbatim}
\usepackage{rotating}
\usepackage{thmtools, thm-restate}
\usepackage{bm}
\usepackage{soul}
\usepackage{hyperref}
\usepackage{mathbbol,physics}
\usepackage{pifont}
\usepackage{appendix}
\usepackage[capitalize]{cleveref}
\usepackage{blkarray}
\usepackage{adjustbox}
\usepackage{graphicx}
\usepackage{dcolumn}

\definecolor{darkred}  {rgb}{0.5,0,0}
\definecolor{darkblue} {rgb}{0,0,0.5}
\definecolor{darkgreen}{rgb}{0,0.5,0}
\hypersetup{
  colorlinks = true,
  urlcolor  = blue,         
  linkcolor = darkblue,     
  citecolor = darkgreen,    
  filecolor = darkred       
}
\theoremstyle{definition}

\usepackage{xcolor}
\definecolor{cool_green}{rgb}{0.0, 0.5, 0.0}

\newcommand{\blk}{\color{black}}

\begin{document}

\title{A contextual advantage for conclusive exclusion: \\
repurposing the Pusey-Barrett-Rudolph construction}

\author{Y{\`i}l{\`e} Y{\=\i}ng}
\email{yying@pitp.ca}
\affiliation{Perimeter Institute for Theoretical Physics, 31 Caroline Street North, Waterloo, Ontario, Canada N2L 2Y5}
\affiliation{Department of Physics \& Astronomy, University of Waterloo, Waterloo, Ontario, Canada N2L 3G1}
\author{David Schmid} 
\affiliation{Perimeter Institute for Theoretical Physics, 31 Caroline Street North, Waterloo, Ontario, Canada N2L 2Y5}
\author{Robert W. Spekkens} 
\affiliation{Perimeter Institute for Theoretical Physics, 31 Caroline Street North, Waterloo, Ontario, Canada N2L 2Y5}
\affiliation{Department of Physics \& Astronomy, University of Waterloo, Waterloo, Ontario, Canada N2L 3G1}

\date{\today}
\begin{abstract}
The task of conclusive exclusion for a set of quantum states is to find a measurement such that for each state in the set, there is an outcome that allows one to conclude with certainty that the state in question was not prepared. Defining classicality of statistics as realizability by a generalized-noncontextual ontological model, we show that there is a quantum-over-classical advantage for how well one can achieve conclusive exclusion. This is achieved in an experimental scenario motivated by the construction appearing in the Pusey-Barrett-Rudolph theorem. We derive noise-robust noncontextuality inequalities bounding the conclusiveness of exclusion, and describe a quantum violation of these. Finally, we show that this bound also constitutes a classical causal compatibility inequality within the bilocality scenario, and that its violation in quantum theory yields a novel possibilistic proof of a quantum-classical gap in that scenario.
\end{abstract}
\maketitle

{\bf Introduction---}
{\em Conclusive exclusion} refers to a quantum state discrimination task where one is given a set $\{\rho_k\}_{k=1}^n$ of quantum states, and one must find a measurement $\{E_k\}_{k=1}^n$ (i.e., with a number of outcomes equal to the number of states in the set) such that obtaining outcome $E_k$ allows one to conclude with certainty that state $\rho_k$ was \emph{not} prepared. The measure of the \emph{conclusiveness of exclusion} can be quantified by 
\begin{align}
   1 - \frac{1}{n} \sum_{k=1}^n \mathbb{P}(E_k|\rho_k),
\end{align}
where $\mathbb{P}(E_k|\rho_k)$ is the probability of obtaining outcome $E_k$ when the state is $\rho_k$. Perfect conclusive exclusion occurs when this quantity is one.

The task of conclusive exclusion was first studied by Caves, Fuchs, and Schack in an article on quantum state compatibility (under the name ``post-Peierls compatibility'')~\cite{CFS2002}. It subsequently became a subject of greater interest when it was used by Pusey, Barrett, and Rudolph (PBR) as part of an argument about the interpretation of the quantum state~\cite{puseyReality2012}.  The term ``conclusive exclusion'' was introduced in Ref.~\cite{PerryJainOppenheim}, where a study was undertaken of what conditions a set of states must satisfy in order for there to be a possibility of conclusive exclusion.  Leifer's review~\cite{leifer2014quantum} of $\psi$-ontology theorems highlighted the significance of  conclusive exclusion (under the name ``antidistinguishability''). Recent work~\cite{Leifer_2020,srikumar2024contextualityantidistinguishabilityrelated} has explored the connection between sets of quantum states for which conclusive exclusion is possible and proofs of the  Kochen-Specker theorem.

We undertake a study of what aspects of  the operational phenomenology of conclusive exclusion resist classical explanation. The notion of classical explainability of experimental statistics we adopt here is realizability by a generalized-noncontextual ontological model~\cite{harriganEinstein2010,Spekkens2005} (henceforth, we use the term “noncontextual” as a shorthand for “generalized-noncontextual"). The arguments in favour of defining classical explainability in this fashion are given in Refs.~\cite{spekkens2019,Spekkens2005,schmid2021guiding,Schmid2021}.  Importantly, this notion of classical explainability is experimentally testable because inequalities for generalized noncontextually are noise-robust.\footnote{Note that the notion of generalized noncontextuality is distinct from that of Kochen-Specker noncontextuality~\cite{Spekkens2005}, and the inequalities derived from Kochen-Specker noncontextuality are not noise-robust~\cite{mazurek2016experimental,kunjwal2015kochen, determinism}. Thus, the aspects of the phenomenology of conclusive exclusion considered in Ref.~\cite{Leifer_2020,srikumar2024contextualityantidistinguishabilityrelated} are not experimentally testable. 
Although there are techniques for converting Kochen-Specker inequalities 
into noise-robust inequalities for generalized noncontextuality~\cite{kunjwal2015kochen,kunjwal2018from}, it is not clear that they can preserve a connection to noise-robust features of conclusive exclusion.  It is an interesting question for future research how to generalize such techniques to do so. 
}
\blk

\begin{figure}[htbp] 
   \centering
   \includegraphics[width=1.5in]{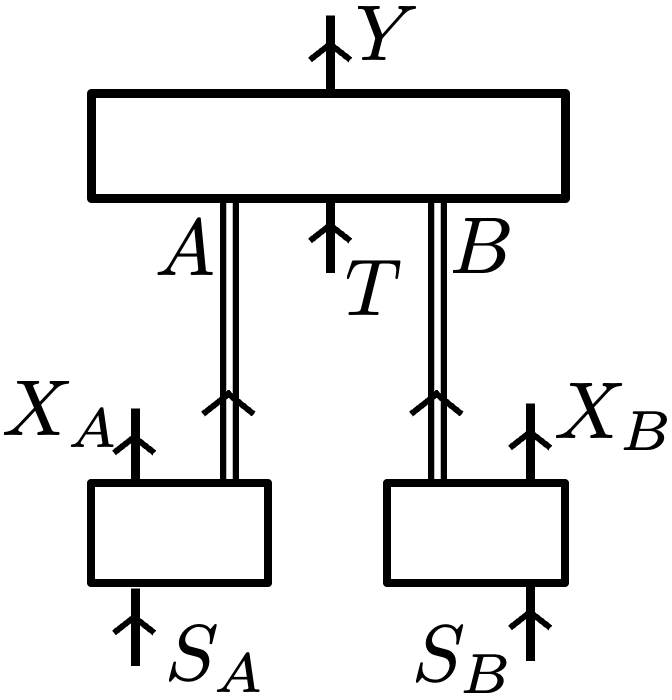} 
   \caption{The quantum circuit diagram for our ${\rm PBR}^+$ scenario. Product states are prepared by a pair of independent multi-sources, and then measured by a joint (entangled) measurement which aims to achieve conclusive exclusion on various quadruples of these product states. Single-stroke wires represent classical systems; double-stroke wires represent quantum systems.}
   \label{fig:setup}
\end{figure}

The causal structure of the experiment considered by PBR involves two systems, $A$ and $B$, that are prepared independently and measured jointly. We consider a slight generalization of this causal structure, depicted in~\cref{fig:setup}, which we refer to as the 
``${\rm PBR}^+$ scenario''\footnote{ Like the original PBR paper, we consider two quantum systems prepared independently and measured jointly; however, we conceptualize the different preparations and measurements via multi-sources and multi-meters.}. 
We show that the quantum prediction for how well one can achieve conclusive exclusion within the ${\rm PBR}^+$ scenario resists classical explanation. 
We do this by showing that such conclusive exclusion implies a novel proof of the impossibility of a noncontextual ontological model. We moreover derive robust noncontextuality inequalities bounding how well one can achieve this task, and we show that quantum theory can violate these bounds. These results help tease out what is genuinely nonclassical about the phenomenology of quantum state discrimination, complementing prior works showing quantum-over-classical advantages for minimum error and unambiguous state discrimination~\cite{schmidContextual2018,statedisc2,statedisc3,statedisc4,statedisc5,statedisc6,flatt2025unifyingcontextualadvantagesstate}. 
They furthermore contribute to the broader project of identifying the precise boundary between classical and quantum for any given phenomenology, as 
has been carried out in the contexts of computation~\cite{Schmid2022Stabilizer,shahandeh2021quantum}, interference~\cite{Catani2023whyinterference,catani2022reply,catani2023aspects}, compatibility~\cite{selby2023incompatibility,selby2023accessible,PhysRevResearch.2.013011}, uncertainty relations~\cite{catani2022nonclassical}, metrology~\cite{contextmetrology}, thermodynamics~\cite{contextmetrology,comar2024contextuality}, weak values~\cite{AWV, KLP19}, coherence~\cite{rossi2023contextuality,Wagner2024coherence}, quantum Darwinism~\cite{baldijao2021noncontextuality}, information processing and communication~\cite{POM,RAC,RAC2,Saha_2019,Yadavalli2020,PhysRevLett.119.220402,fonseca2024robustness}, cloning~\cite{cloningcontext}, broadcasting~\cite{jokinen2024nobroadcasting}, as well as Bell scenarios~\cite{Wright2023invertible,schmid2020unscrambling} and Kochen-Specker scenarios~\cite{operationalks,kunjwal2018from,Kunjwal16,Kunjwal19,Kunjwal20,specker,Gonda2018almostquantum}.

Finally, we show that the existence of noncontextual inequalities for the ${\rm PBR}^+$ scenario implies the existence of classical causal compatibility inequalities for the bilocality scenario~\cite{bilocality}.  That is, we demonstrate a bridge between the two sorts of scenarios and no-go theorems and inequalities pertaining to each. Via this bridge, we establish that conclusive exclusion is implicated in a quantum violation of a classical causal compatibility inequality in the bilocality scenario. Thus our work also contributes to the broader project of identifying networks with classical causal incompatibility inequalities that can be violated by quantum theory~\cite{tavakoli2021bell,bilocality,fritz2012beyond,wood2015lesson,Chaves2015InfoTheo,Wolfe2016inflation,Chaves2018,Himbeeck2018instrumental,wolfe2019quantum,TriangleGisin,triangle,weilenmann2017analysing,krivachy2020neural,PhysRevLett.128.060401,PhysRevLett.116.010402}.

{\bf Conclusive exclusion in the ${\rm PBR}^+$ scenario---}
Our construction is inspired by the particular conclusive exclusion task that is used in the proof of the PBR theorem. (This operational task should not be conflated with the PBR theorem itself, which is sometimes believed to be a strong argument against $\psi$-epistemic ontological models of operational quantum theory---a conclusion we reject for the reasons described in Ref.~\cite{SpePIRSA}.) 

We take the two systems in the ${\rm PBR}^+$ scenario to be two qubits and denote them $A$ and $B$. We now define the construction, beginning with a particular choice of product states on the composite $AB$.
The basic construction presumes that each qubit has a state chosen from the set $\Omega_{0+} \coloneqq \{|0\rangle, |+\rangle \}$
so that the possible product states for the two qubits form the set denoted $\mathcal{P}_{0+}\coloneqq\{\rho_{k|0+}\}_{k=1}^4$ with (nonorthogonal) elements
\begin{subequations}\label{rho0+}
\begin{align}
   \rho_{1|0+}\coloneqq&\ketbra{0}_A \otimes \ketbra{0}_B,\label{PBRstate1}\\
\rho_{2|0+}\coloneqq&\ketbra{0}_A \otimes \ketbra{+}_B,\label{PBRstate2}\\
\rho_{3|0+}\coloneqq&\ketbra{+}_A \otimes \ketbra{0}_B,\label{PBRstate3}\\
\rho_{4|0+}\coloneqq&\ketbra{+}_A \otimes \ketbra{+}_B\label{PBRstate4}. 
\end{align}
\end{subequations}
(Here, the notation $\mathcal{P}$ is chosen to remind the reader it is a set of product states.)
The measurement that achieves the conclusive exclusion of the states in this set is the one associated to the orthonormal basis $\mathcal{M}_{0+} \coloneqq \{E_{k|0+}\coloneqq \ketbra{\phi_{k|0+}}
\}_{k=1}^4$, where
\begin{subequations}
\begin{align}
\label{PBRmeasurement}
\ket{\phi_{1|0+}}\coloneqq&
\frac{1}{\sqrt{2}} \left( |0 \rangle_A |1\rangle_B + |1 \rangle_A |0\rangle_B \right), \\
\ket{\phi_{2|0+}}\coloneqq&
\frac{1}{\sqrt{2}} \left( |0 \rangle_A |-\rangle_B + |1 \rangle_A |+\rangle_B \right),\\
\ket{\phi_{3|0+}}\coloneqq&
\frac{1}{\sqrt{2}} \left( |+ \rangle_A |1\rangle_B + |- \rangle_A |0\rangle_B \right),\\
\ket{\phi_{4|0+}}\coloneqq&
\frac{1}{\sqrt{2}} \left( |+ \rangle_A |-\rangle_B + |- \rangle_A |+\rangle_B \right).
\end{align}
\end{subequations}
This is easily verified to achieve conclusive exclusion because $E_{1|0+}$ is orthogonal to $\rho_{1|0+}$, $E_{2|0+}$ is orthogonal to $\rho_{2|0+}$, and so on.

This result about conclusive exclusion is invariant under application to the state of any unitary that factorizes across the $AB$ partition, since applying the inverse of this unitary to the measurement maintains the operational statistics.
Thus, we can introduce rotated sets of states for which one can also achieve perfect conclusive exclusion using inversely rotated measurements. 

In this way, we will consider four related conclusive exclusion protocols.

In particular, we consider three other sets of pairs of single qubit states $\Omega_{0-} \coloneqq \{|0\rangle, |-\rangle \},$ $\Omega_{1+} \coloneqq \{|1\rangle, |+\rangle \},$ and $\Omega_{1-} \coloneqq \{|1\rangle, |-\rangle \}$, for each of which we consider the set of nonorthogonal product states on $AB$ that they define. The sets are respectively denoted $\mathcal{P}_{0-}\coloneqq \{ \rho_{k|0-}\}_{k=1}^4$, $ \mathcal{P}_{1+}\coloneqq \{ \rho_{k|1+}\}_{k=1}^4$ and $ \mathcal{P}_{1-}\coloneqq \{ \rho_{k|1-}\}_{k=1}^4$, where
\begin{subequations}\label{otherrho}
\begin{align}
    \rho_{k|0-}\coloneqq (Z \otimes Z) \rho_{k|0+} (Z \otimes Z),\\
   \rho_{k|1+}\coloneqq (X \otimes X) \rho_{k|0+} (X \otimes X),\\
    \rho_{k|1-}\coloneqq (Y \otimes Y) \rho_{k|0+} (Y \otimes Y).
\end{align}
\end{subequations}
The corresponding measurements that achieve conclusive exclusion are $\mathcal{M}_{0-}\coloneqq \{ E_{k|0-}\}_{k=1}^4$, $ \mathcal{M}_{1+}\coloneqq \{ E_{k|1+}\}_{k=1}^4$ and $ \mathcal{M}_{1-}\coloneqq \{ E_{k|1-}\}_{k=1}^4$, where
\begin{subequations}
\begin{align}
E_{k|0-}\coloneqq (Z \otimes Z) E_{k|0+} (Z \otimes Z),\\
E_{k|1+}\coloneqq (X \otimes X) E_{k|0+} (X \otimes X),\\
E_{k|1-}\coloneqq (Y \otimes Y) E_{k|0+} (Y \otimes Y).
\end{align}
\end{subequations}
Define a $Z$-source as a preparation procedure that
randomly prepares one of the states in $\{|0\rangle, |1\rangle\}$, and outputs the bit $0$ in the $|0\rangle$ case and outputs the bit $1$ in the $|1\rangle$ case. Similarly define an $X$-source as a preparation procedure that randomly prepares one of the states in $\{|+\rangle, |-\rangle\}$, and outputs the bit $0$ in the $|+\rangle$ case and outputs the bit $1$ in the $|-\rangle$ case. Next consider a device that toggles between the $Z$-source and $X$-source based on the value of a binary setting variable. This is an instance of a {\em multi-source}~\cite{zhang2025reassessingboundaryclassicalnonclassical}.  
The experiment in Fig.~\ref{fig:setup} involves a multi-source $\mathcal{S}^A$ on qubit $A$ with setting variable $S_A$ and outcome variable $X_A$, and a multi-source $\mathcal{S}^B$ on qubit $B$ with setting variable $S_B$ and outcome variable $X_B$. 

\begin{figure}[htbp] 
    \centering
    \includegraphics[width=3.45in]{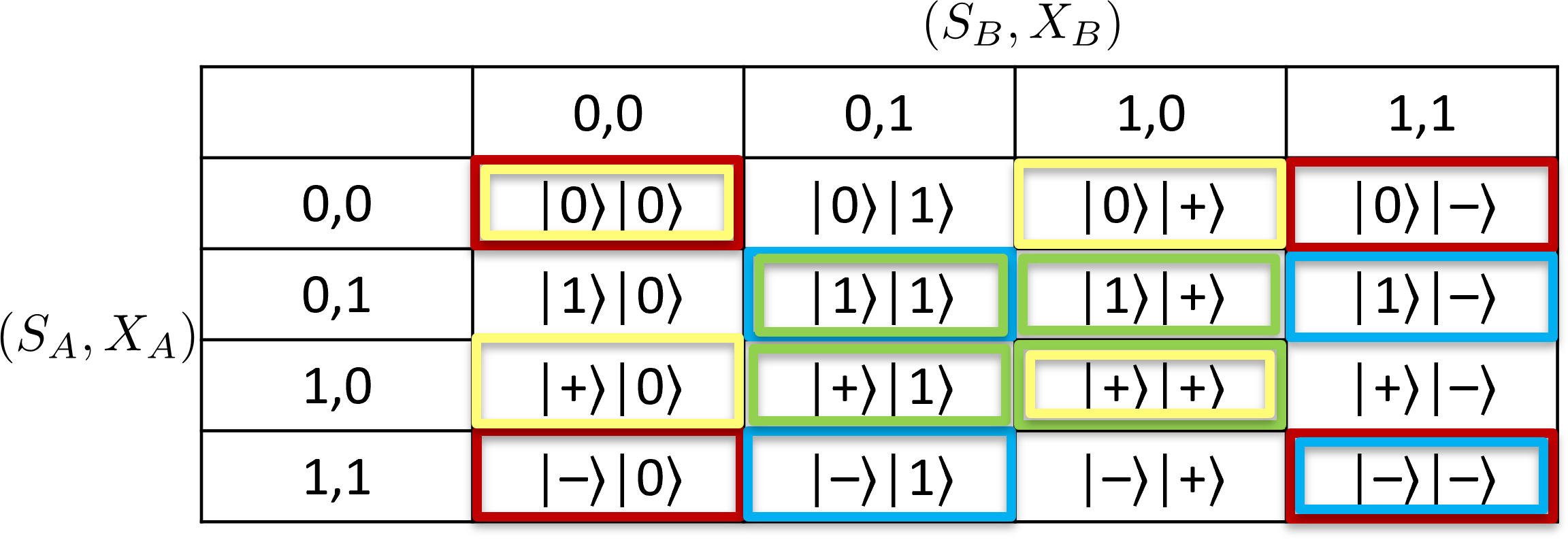} 
    \caption{The product states generated by the pair of multi-sources. Quadruples of states which can be consistently excluded are circled in matching colors. Yellow  (states in ${\cal P}_{0+}$): $(S_A,X_A),(S_B,X_B) \in \{ (0,0), (1,0) \} \times \{ (0,0), (1,0) \}$. Red  (states in ${\cal P}_{0-}$): $(S_A,X_A),(S_B,X_B) \in \{ (0,0), (1,1) \} \times \{ (0,0), (1,1) \}$. Green  (states in ${\cal P}_{1+}$): $(S_A,X_A),(S_B,X_B) \in \{ (0,1), (1,0) \} \times \{ (0,1), (1,0) \}$. Blue  (states in ${\cal P}_{1-}$): $(S_A,X_A),(S_B,X_B) \in \{ (0,1), (1,1) \} \times \{ (0,1), (1,1) \}$.} 
    \label{fig:table}
 \end{figure}

The 16 possible product states on $AB$ that this pair of multi-sources can generate,  as a function of the different possible values of $(S_A,X_A;S_B,X_B)$, are listed in \cref{fig:table}.  All of the states in our four conclusive exclusion protocols can be so generated, and so appear in the table. (Note that four of the product states in the table do not appear in any of our conclusive exclusion protocols, while the four along the diagonal of the table each appear in {\em two} of the  conclusive exclusion protocols.) 
The states appearing in $\mathcal{P}_{0+}$
are highlighted in yellow, 
those in $\mathcal{P}_{0-}$ in red, those in $\mathcal{P}_{1+}$ in green, and those in $\mathcal{P}_{1-}$ are highlighted in blue. 
 
The multi-measurement implemented on $AB$ has a setting variable $T\in \{ 0+,0-,1+,1-\}$ that toggles among the four measurements $\mathcal{M}_{T}$ that achieve conclusive exclusion for each corresponding set $\mathcal{P}_{T}$.  Specifically, the outcome $Y\in\{1,2,3,4\}$ of $\mathcal{M}_{T}$, namely, the one corresponding to $E_{Y|T}$, conclusively excludes the $Y$th state in $\mathcal{P}_{T}$, namely, the state $\rho_{Y|T}$. 

The measure of conclusiveness of exclusion for each of the tasks is
\begin{align}
\label{eq:CE}
    {\rm CE}_{T} \coloneqq 1 - \frac{1}{4} \sum_Y \mathbb{P}(E_{Y|T}|\rho_{Y|T}).
\end{align}
 
${\rm CE}_T$ can equivalently be written more explicitly in terms of the classical variables  $Y$, $T$, $S_A$, $X_A$, $S_B$ and $X_B$ appearing in Figure~\ref{fig:table},  (instead of only $Y$ and $T$), as 
\begin{align}
\label{eq:CESX}
    {\rm CE}_{T} = 1 - \frac{1}{4} \sum_Y \mathbb{P}(E_{Y|T}|\tilde{\rho}_{S_A(Y,T),X_A(Y,T),S_B(Y,T),X_B(Y,T)}),
\end{align}
where $S_A(Y,T)$ indicates the value of $S_A$ as a function of the tuple $(Y,T)$, and similarly for $X_A(Y,T)$, $S_B(Y,T)$ and $X_B(Y,T)$, such that $\tilde{\rho}_{S_A(Y,T),X_A(Y,T),S_B(Y,T),X_B(Y,T)}=\rho_{Y|T}$. 
(The explicit form of these functions can be identified by checking how the $\rho_{Y|T}$s defined in \cref{rho0+,otherrho} correspond to states generated by $(S_A, X_A, S_B, X_B)$ as shown in \cref{fig:table}.)

Averaging the conclusiveness of exclusion for the four tasks, we have the quantity for which we will provide a noncontextual bound, namely, 
 \begin{align}
 {\rm CE} &\coloneqq  \frac{1}{4} ( {\rm CE}_{0+} +{\rm CE}_{0-} +{\rm CE}_{1+}+{\rm CE}_{1-} ).
 \end{align}

It will be important in what follows that the qubit states $\{ |0\rangle, |1\rangle, |+\rangle, |-\rangle\}$ satisfy the operational identity
\begin{equation}\label{opeq}
\frac{1}{2} |0\rangle\langle 0| +\frac{1}{2} |1\rangle\langle 1| = \frac{1}{2} |+\rangle\langle +| +\frac{1}{2} |-\rangle\langle -|,
\end{equation}
expressing the fact that the $Z$-source and $X$-source produce the same average quantum state if one marginalizes over their outcomes.

{\bf Noncontextual bound on the success rate---}
The operational identity of \cref{opeq}, together with the assumption of noncontextuality, implies constraints on the conclusiveness of exclusion. To prove this, we use the fact~\cite{Schmid2021,Schmid2024structuretheorem} that if we view quantum theory as a generalized probabilistic theory~\cite{Hardy,Barrett2006,chiribella2010probabilistic} (quotiented with respect to contexts~\cite{dariano2017quantum,Schmid2024structuretheorem,schmid2020unscrambling}), consistency with the principle of noncontextuality is equivalent to the existence of a linear representation that is positive for all states and effects. Such a model for the scenario of Fig.~\ref{fig:setup} must have the following form. One associates a set of ontic states to system $A$, denoted $\Lambda_A$,  and a set of ontic states to system $B$, denoted $\Lambda_B$. One then associates to each quantum state on system $A$ a probability distribution over this set, denoted $\mu_\rho(\Lambda_A)$, which is a linear function of the density matrix. Similarly for system $B$. One associates to each POVM element a response function, denoted $\xi(y|\lambda_A,\lambda_B)$, which is a linear function of the POVM element.
Finally, the quantum predictions must be reproduced via
\begin{align}
    &\tr[E_y^{AB} (\rho_A \otimes \rho_B)] \\
    =&\int_{\Lambda_A \times \Lambda_B} \xi(y|\lambda_A,\lambda_B) \mu_{\rho_A}(\lambda_A) \mu_{\rho_B}(\lambda_B) d \lambda_A d\lambda_B. \nonumber
\end{align}

Let $\mu_0(\lambda), \mu_1(\lambda), \mu_+(\lambda),$ and $\mu_-(\lambda)$ denote the distributions associated to the quantum states $|0\rangle,|1\rangle, |+\rangle$, and $|-\rangle$ respectively. From the operational identity of Eq.~\eqref{opeq}, the linear dependence of the distributions on quantum states implies that
\begin{equation}\label{averagesourcePNC}
\frac{1}{2} \mu_0(\lambda) +\frac{1}{2} \mu_1(\lambda) = \frac{1}{2} \mu_+(\lambda) +\frac{1}{2} \mu_-(\lambda).
\end{equation}

Let $\tilde{\Omega}^A_{T}$ denote the set of distributions that represent the set of quantum states $\Omega^A_{T}$, and $\tilde{\Omega}^B_{T}$ the set of distributions that represent the set $\Omega^B_{T}$.

\begin{figure}[h!] 
\centering
\includegraphics[width=0.5\textwidth]{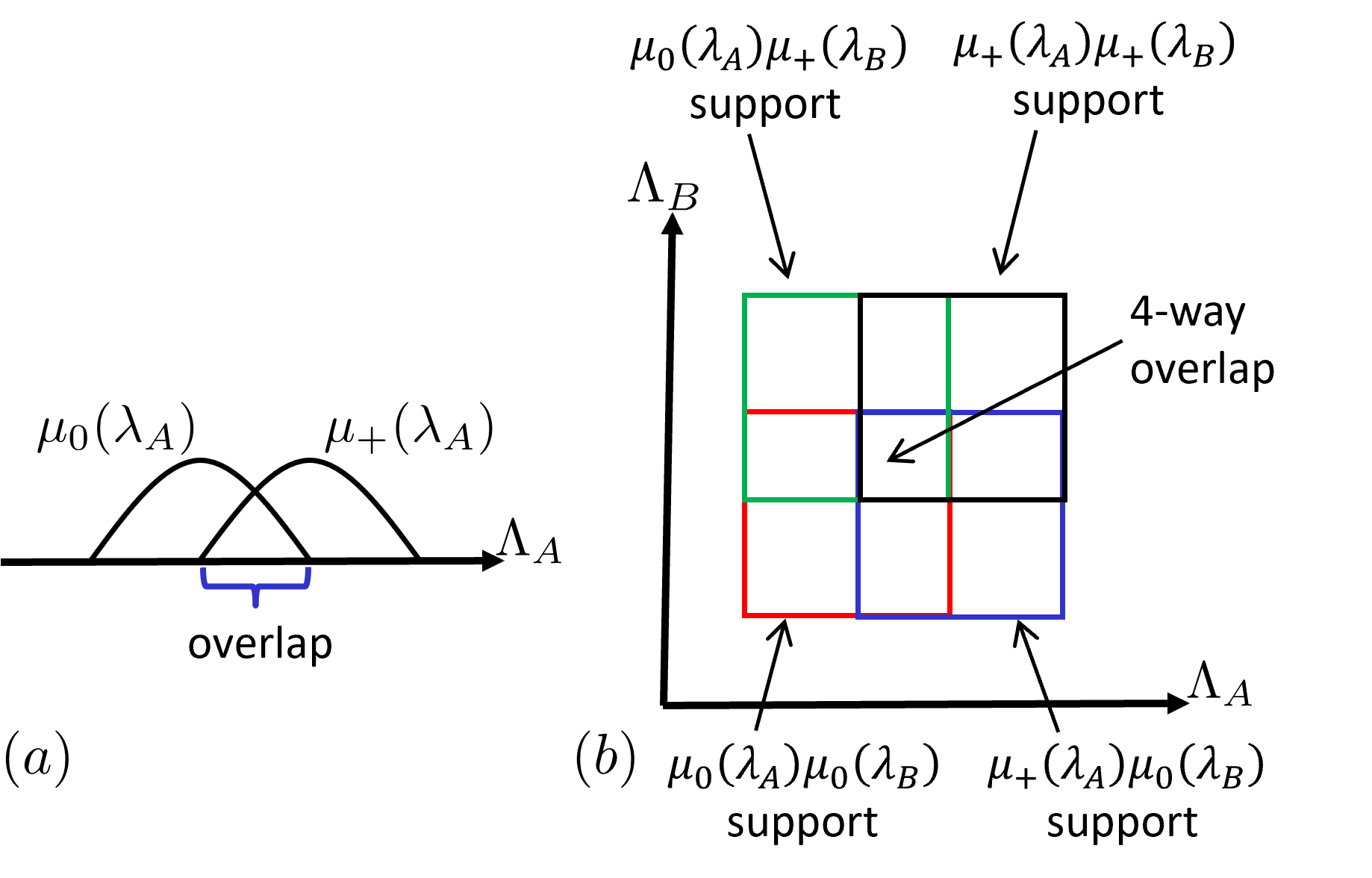}
\caption{(a) A depiction of the probability distributions over $\Lambda_A$ for the $\ket{0}$ and $\ket{+}$ states having overlapping ontic supports.  (b) A depiction of the ontic supports of the probability distributions over $\Lambda_A\times \Lambda_B$ for the four product states $\ket{0}\ket{0}$, $\ket{0}\ket{+}$, $\ket{+}\ket{0}$, and $\ket{+}\ket{+}$. Assuming the supports of $\ket{0}$ and $\ket{+}$ overlap, as in (a), it follows that there is a region where all four overlap. 
} 
\label{PBROnticSupportsForPair}
\end{figure}

As shorthand, we say that a set $\Omega^A_{T}$ of quantum states {\em has nontrivial ontic overlap} if the elements of the corresponding set $\tilde{\Omega}^A_{T}$ of probability distributions on $\Lambda_A$ have a nontrivial overlap of their ontic supports.  
For instance, if the set of quantum states $\Omega^A_{0+}  = \{ |0\rangle^A, |+\rangle^A\}$ is represented by the set of epistemic states $\tilde{\Omega}^A_{0+} = \{ \mu_0(\lambda_A),\mu_+(\lambda_A)\}$ of the sort depicted in Fig.~\ref{PBROnticSupportsForPair}(a), then this set of quantum states is said to have nontrivial ontic overlap. We make an analogous definition for sets of states on $B$ and on the composite $AB$.

Under the assumption of noncontextuality, we prove in Appendix~\ref{app:overlap} that $\ket{0}$ and $\ket{+}$ must have nontrivial ontic overlap. The same is true for $\ket{0}$ and $\ket{-}$, for $\ket{1}$ and $\ket{+}$, and for $\ket{1}$ and $\ket{-}$.\footnote{This strengthens a similar claim in Ref.~\cite{harriganEinstein2010}, namely, that the principle of noncontextuality can only be satisfied in ontological models that are $\psi$-epistemic, and more precisely, at least one of the four pairs must have nontrivial ontic overlap.} 

If the 2-element set $\Omega^A_{0+}$ has nontrivial ontic overlap then so does the 2-element set $\Omega^B_{0+}$ and so does the set of product states $\mathcal{P}_{0+}$. The latter implication can be understood by noting that the ontic support of a product state in $\mathcal{P}_{0+}$ is simply the Cartesian product of the ontic supports of its factors. This is illustrated in  Fig.~\ref{PBROnticSupportsForPair}.  

We noted above that each of the four sets of states, $\Omega^A_{0+}$, $\Omega^A_{0-}$, $\Omega^A_{1+}$ and $\Omega^A_{1-}$ has nontrivial ontic overlap.  Therefore, each of the four sets of product states,  $\mathcal{P}_{0+}$, $\mathcal{P}_{0-}$, $\mathcal{P}_{1+}$ and $\mathcal{P}_{1-}$, must also have nontrivial ontic overlap.  
 
But any set of product states that has nontrivial ontic overlap is one for which perfect conclusive exclusion is impossible.  This is because even in a theory allowing a perfect measurement that reveals the exact ontic state of the system, ontic states in the region of overlap are consistent with having been sampled from any of the distributions, and so none of the distributions can be excluded conclusively.

So, in any model consistent with  noncontextuality, the average of the success rates for conclusive exclusion for the four tasks, namely ${\rm CE}$,  must be bounded away from 1.
But quantum theory predicts that ${\rm CE} =1$, since the optimal quantum conclusive exclusion in each of these four exclusion tasks above is 1. Consequently, the (perfect) conclusive exclusion allowed by quantum theory is a proof of the nonrealizability of the statistics by a noncontextual ontological model, hence of nonclassicality.

In Appendix~\ref{app:proofCE}, we determine the precise amount by which CE is bounded away from $1$ in any noncontextual model, namely
\begin{equation}
{\rm CE}  \leq_{\text{nc}} \frac{15}{16} = 93.75\%
\label{NCinequality}
\end{equation}
This is a noise-robust noncontextuality inequality~\cite{Mazurek2016,schmidAll2018}, expressed in terms of the average of four conclusiveness of exclusion quantities.  Having a noise-robust inequality is important for achieving an experimental test, since any real-world experiment will have noise and imperfections that cause CE to be bounded away from its logical maximum of 1.  However, as long as the noise and imperfections are small enough, one can still violate the noncontextual bound.

\begin{figure}[h!] 
\centering
\includegraphics[width=0.3\textwidth]{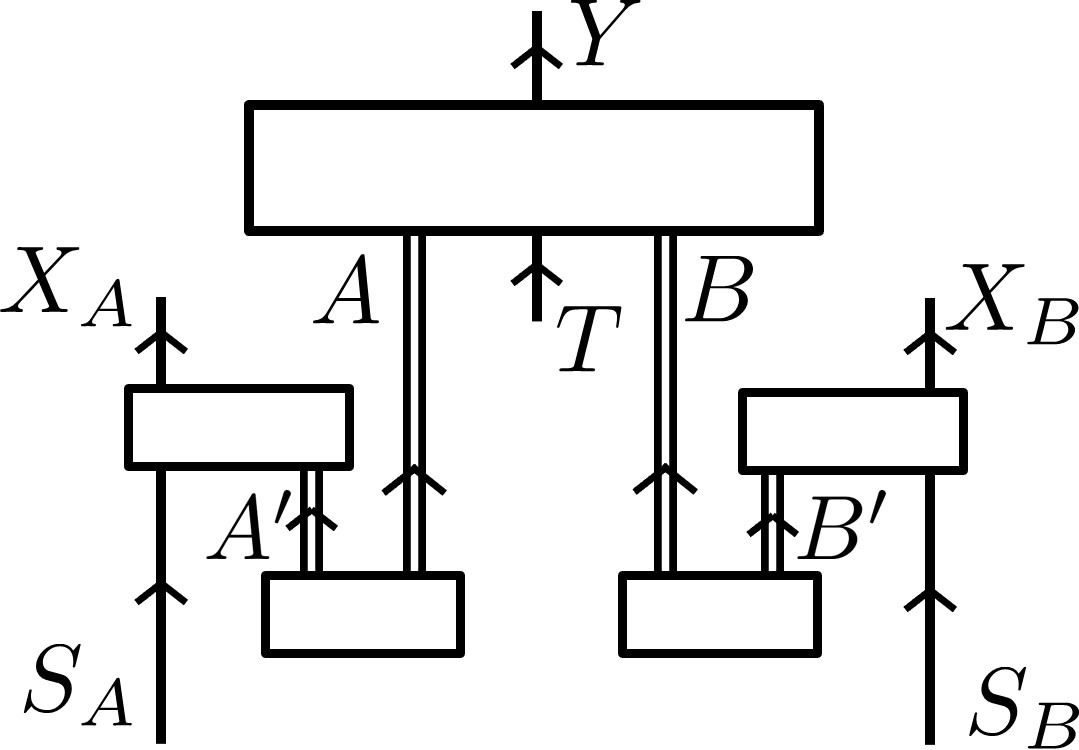}
\caption{The circuit diagram for the bilocality scenario.   } 
\label{BilocalityScenario}
\end{figure}

{\bf Connection to the bilocality scenario---} 
Assumptions about the {\em causal structure} that holds among a number of variables 
can imply constraints on the correlations that can be realized on these variables, with the paradigm example being how the Bell causal structure (with a classical latent variable as a common cause) implies that the distribution over outcome variables given setting variables should satisfy Bell inequalities. 
Generally, inequality constraints on the realizable correlations when the latent nodes are assumed to be classical in a given causal structure are  termed {\em classical causal compatibility inequalities} for that causal structure~\cite{Wolfe2016inflation} (see Refs.~\cite{pearl2009causality,fritz2012beyond,Himbeeck2018instrumental} for examples). In Appendix~\ref{app:bilocality}, we show that the noncontextuality inequality of \cref{NCinequality} for the ${\rm PBR}^+$ scenario (Fig.~\ref{fig:setup}) 
arises as a classical causal compatibility inequality within the causal structure known as the bilocality scenario~\cite{bilocality}, depicted in Fig.~\ref{BilocalityScenario}. (Recall that from Eq.~\eqref{eq:CESX}, this inequality can be expressed in terms of the conditional probabilities $\mathbb{P}(X_A X_B Y| S_A S_B T)$ that are native to the bilocality scenario.) That is, if one is certain that the causal structure of one's experiment is as shown in Fig.~\ref{BilocalityScenario}, then one does not even need to appeal to noncontextuality to derive the bound in Eq.~\eqref{NCinequality}---rather, the bound follows from the causal structure itself. 

One way to obtain a nontrivial classical causal compatibility inequality (i.e., one that can be violated by quantum theory) in the bilocality scenario is to piggyback on Bell's theorem: by postselecting on the joint measurement's outcome, one can achieve entanglement swapping, effectively reducing the scenario to a standard Bell scenario with entanglement between $A'$ and $B'$.   It follows that one can achieve every variety of proof that is achievable in the standard Bell scenario, including {\em possibilistic} proofs (the paradigm example of which is Hardy's proof~\cite{Hardy1993paradox}) wherein the quantum-classical gap is manifest already at the level of the {\em support} of the distribution, i.e., a specification of which outcomes are possible and which are impossible, without regard to the values of the nonzero probabilities. However, the question of interest has always been whether there are {\em novel} proofs of quantum-classical gaps in the bilocality scenario, i.e., proofs which do not piggyback on Bell's theorem through the entanglement swapping protocol~\cite{bilocality,BilocalCorrelations,
ciudadalanon2024escapingshadowbellstheorem}. The quantum-classical gap we have manifested here is of this variety. Furthermore, to our knowledge, it is the first example of a novel proof of nonclassicality in the bilocality scenario that is {\em possibilistic}.

{\em Acknowledgements---}
DS, RWS, and YY were supported by Perimeter Institute for Theoretical Physics. Research at Perimeter Institute is supported in part by the Government of Canada through the Department of Innovation, Science and Economic Development and by the Province of Ontario through the Ministry of Colleges and Universities. YY was also supported by the Natural Sciences and Engineering Research Council of Canada (Grant No. RGPIN-2024-04419).

\bibliography{PBR.bib} 

\appendix

  \begin{widetext}
  
\section{Overlapping ontic supports}
\label{app:overlap}

Here we prove that each of the four sets of states, $\Omega^A_{0+}$, $\Omega^A_{0-}$, $\Omega^A_{1+}$ and $\Omega^A_{1-}$, must have nontrivial ontic overlap.

\begin{proof}
Denote the respective ontic supports for $\mu_0(\lambda),\mu_1(\lambda),\mu_+(\lambda), \mu_-(\lambda)$ as $\Lambda_0,\Lambda_1,\Lambda_+,\Lambda_-$.

As mentioned in the main text, under the assumption of linear representations of states and effects in the ontological model of the quotiented theory (which is equivalent to the assumption of noncontextuality for the unquotiented theory), the operational identity among these four states, namely, \cref{opeq}, implies the corresponding identity among their ontic representations, namely, \cref{averagesourcePNC}.

As proven in Ref.~\cite{schmidContextual2018}, $\Lambda_0\neq\Lambda_+$, thus, there exists $ \lambda^{\ast}\in\Lambda_0$ such that $\lambda^{\ast}\notin  \Lambda_{+}$, i.e., $\mu_0(\lambda^{\ast})>0$ while $\mu_+(\lambda^{\ast})=0$. Then, together with the fact that $\mu_1(\lambda)$ is nonnegative, \cref{averagesourcePNC} implies that $\mu_-(\lambda^{\ast})>0$. Thus, $\lambda^{\ast}$ is in the ontic support for both $\mu_0$ and $\mu_-$, implying that $\Omega_{0-}$ has nontrivial ontic overlap.

Similarly, since it is proven in Ref.~\cite{schmidContextual2018} that $\Lambda_0\neq\Lambda_-$, $\Lambda_1\neq\Lambda_+$, and $\Lambda_1\neq\Lambda_-$, the analogue of the argument just presented implies that $\Omega_{0+}$, $\Omega_{1-}$ and $\Omega_{1+}$ have nontrivial ontic overlap.

\end{proof}

\section{Proof of the noncontextuality inequality}
\label{app:proofCE}

Here we derive the noncontextuality inequality on CE, namely, \cref{NCinequality}, and show that it is tight, meaning that the maximal value of CE that can be achieved for any theory that admits of a noncontextual ontological model is 15/16. Such a proof does not assume the correctness of quantum theory, so we first translate the scenario described in the main text to one for a general GPT.

In the experimental scenario depicted in \cref{fig:setup}, the $Z$-source now toggles between the GPT states $\vec{s}_0$ and $\vec{s}_1$ while the $X$-source now toggles between the GPT states $\vec{s}_+$ and $\vec{s}_-$ such that they satisfy the operational identity:
\begin{align}
\label{eq:GPTopid}
    \frac{1}{2}\vec{s}_0+\frac{1}{2}\vec{s}_1=\frac{1}{2}\vec{s}_++\frac{1}{2}\vec{s}_-.
\end{align}
We use $\Omega_{0+}$ to denote the set $\{\vec{s}_0, \vec{s}_+\}$, and use $\mathcal{P}_{0+}$ for the set of product states generated by it, namely, $\{\vec{s}_{1|0+}\coloneqq\vec{s}_0\otimes \vec{s}_0, \vec{s}_{2|0+}\coloneqq\vec{s}_0\otimes \vec{s}_+,\vec{s}_{3|0+}\coloneqq\vec{s}_+\otimes \vec{s}_0,\vec{s}_{4|0+}\coloneqq\vec{s}_+\otimes \vec{s}_+\}$. Similarly, $\Omega_{0-}\coloneqq \{\vec{s}_0, \vec{s}_-\}$, $\Omega_{1+} \coloneqq \{\vec{s}_1, \vec{s}_+ \},$ and $\Omega_{1-} \coloneqq \{\vec{s}_1, \vec{s}_-\}$, and the corresponding sets of product states are denoted as $\mathcal{P}_{0-}\coloneqq \{\vec{s}_{k|0-}\}_k$, $\mathcal{P}_{1+}\coloneqq \{\vec{s}_{k|1+}\}_k$ and $\mathcal{P}_{1-}\coloneqq \{\vec{s}_{k|1-}\}_k$.

The measurement now toggles between four GPT measurements, each of them is labeled by the measurement setting $T\in\{0+,0-,1+,1-\}$ and have four outcomes labeled by $Y\in\{1,2,3,4\}$. For example, $\mathcal{M}_{0+}\coloneqq\{\vec{e}_{1|0+},\vec{e}_{2|0+},\vec{e}_{3|0+},\vec{e}_{4|0+}\}$.

The optimal conclusiveness of exclusion that can be achieved by the set of states $\mathcal{P}_{T}$ and the measurement $\mathcal{M}_{T}$ is now 
\begin{align}
    {\rm CE}_{T} \coloneqq 1- \frac{1}{4}\sum^{4}_{Y=1} \mathbb{P}(\vec{e}_{Y|T}|\vec{s}_{Y|T}) = 1- \frac{1}{4}\sum^{4}_{Y=1} \vec{e}_{Y|T}\cdot \vec{s}_{Y|T}. \\ \nonumber
\end{align}
Summing the conclusiveness of exclusion for the four tasks, we have the quantity for which we will provide a noncontextual bound, namely, 
 \begin{align}
 {\rm CE} &\coloneqq \frac{1}{4} ( {\rm CE}_{0+} +{\rm CE}_{0-} +{\rm CE}_{1+}+{\rm CE}_{1-} ).
 \end{align}

The ontic state space is denoted $\Lambda$. The ontic representations of $\vec{s}_0, \vec{s}_1, \vec{s}_+, \vec{s}_-$ are denoted by 
$\mu_0(\lambda),\mu_1(\lambda),\mu_+(\lambda), \mu_-(\lambda)$, respectively; and their corresponding ontic supports are denoted  $\Lambda_0,\Lambda_1,\Lambda_+,\Lambda_-$. The operational identity of \cref{eq:GPTopid} together with the assumption of noncontextuality implies 
\begin{align}\label{onid}
    \frac{1}{2}\mu_0(\lambda)+\frac{1}{2}\mu_1(\lambda)=\frac{1}{2}\mu_+(\lambda)+\frac{1}{2}\mu_-(\lambda).
\end{align}

The ontic state space of a bipartite system is simply the Cartesian product, namely, $\Lambda_A \times \Lambda_B =\Lambda \times \Lambda$, and the ontic representation of a product state is the product of the ontic representations of the unipartite state in the product. We denote the response function representing the measurement effect $\vec{e}_{Y|T}$ as $\xi_{Y|T}(\lambda_A,\lambda_B)$ and denote the probability distribution representing the GPT state on $AB$, $\vec{s}_{Y|T}$, as  $\mu_{Y|T}(\lambda_A,\lambda_B)$, such that
\begin{equation}
    \mathbb{P}(\vec{e}_{Y|T}|\vec{s}_{Y|T})=\int_{\Lambda} \int_{\Lambda}\xi_{Y|T}(\lambda_A,\lambda_B)\mu_{Y|T}(\lambda_A,\lambda_B)d\lambda_A d\lambda_B.
\end{equation}

Let us first consider ${\rm CE}_{0+}$. We have \begin{align}
    {\rm CE}_{0+}=&1-\frac{1}{4}\sum^{4}_{Y=1}  \int_{\Lambda} \int_{\Lambda}  \xi_{Y|0+}(\lambda_A,\lambda_B)\mu_{Y|0+}(\lambda_A,\lambda_B)d\lambda_A d\lambda_B. \\ \nonumber
\end{align}

Since $(\Lambda_{0}\cap\Lambda_{+}) \subseteq \Lambda$, we have
\begin{align}
    {\rm CE}_{0+} 
    \leq & 1- \frac{1}{4}\sum^{4}_{Y=1} \int_{\Lambda_{0}\cap\Lambda_{+}}\int_{\Lambda_{0}\cap\Lambda_{+}}\xi_{Y|0+}(\lambda_A,\lambda_B)\mu_{Y|0+}(\lambda_A,\lambda_B)d\lambda_A d\lambda_B.
\end{align}
Expanding the probability distributions on the composite $AB$ into the probability distributions on the components $A$ and $B$, we have
\begin{align}
\nonumber 
    {\rm CE}_{0+}
    \leq &1-\frac{1}{4}\int_{\Lambda_{0}\cap\Lambda_{+}} \int_{\Lambda_{0}\cap\Lambda_{+}}  [\xi_{1|0+}(\lambda_A,\lambda_B)\mu_{0}(\lambda_A)\mu_{0}(\lambda_B)+ \xi_{2|0+}(\lambda_A,\lambda_B)\mu_{0}(\lambda_A)\mu_{+}(\lambda_B)   \\
    &+ \xi_{3|0+}(\lambda_A,\lambda_B)\mu_{+}(\lambda_A)\mu_{0}(\lambda_B) +
    \xi_{4|0+}(\lambda_A,\lambda_B)\mu_{+}(\lambda_A)\mu_{+}(\lambda_B) ]
    d\lambda_A d\lambda_B \nonumber \\ 
    \leq& 1- \frac{1}{4}\int_{\Lambda_{0}\cap\Lambda_{+}}\int_{\Lambda_{0}\cap\Lambda_{+}} \sum^{4}_{Y=1}\xi_{Y|0+}(\lambda_A,\lambda_{B})
    \min\{\mu_{0}(\lambda_A),\mu_{+}(\lambda_A)\} \min\{\mu_{0}(\lambda_B),\mu_{+}(\lambda_B)\} d\lambda_A d\lambda_B \nonumber \\ 
    =&1- \frac{1}{4}\left[\int_{\Lambda_{0}\cap\Lambda_{+}} 
    \min\{\mu_{0}(\lambda),\mu_{+}(\lambda) \} d\lambda \right]^2.
\end{align}
We can repeat the above analysis for the other three terms in CE and get:
\begin{align}
\label{eq:each}
    {\rm CE}_{0+} \leq 1 - \frac{1}{4}
    \left[\int_{\Lambda_{0}\cap\Lambda_{+}} \min\{\mu_{0}(\lambda),\mu_{+}(\lambda) \} d\lambda \right]^2, \\ 
    {\rm CE}_{0-} \leq 1 -  \frac{1}{4}\left[\int_{\Lambda_{0}\cap\Lambda_{-}} \min\{\mu_{0}(\lambda),\mu_{-}(\lambda) \} d\lambda \right]^2, \\ 
    {\rm CE}_{1+} \leq 1 - \frac{1}{4}\left[\int_{\Lambda_{1}\cap\Lambda_{+}} \min\{\mu_{1}(\lambda),\mu_{+}(\lambda) \} d\lambda \right]^2, \\ 
    {\rm CE}_{1-} \leq 1 - \frac{1}{4} \left[\int_{\Lambda_{1}\cap\Lambda_{-}} \min\{\mu_{1}(\lambda),\mu_{-}(\lambda) \} d\lambda \right]^2.
\end{align}

Thus, 
\begin{align}
\label{eq:generaloptimal}
    {\rm CE} \leq 1 - \frac{1}{16} \left\{
    \left[\int_{\Lambda_{0}\cap\Lambda_{+}} \min\{\mu_{0}(\lambda),\mu_{+}(\lambda) \} d\lambda \right]^2 
    + \left[\int_{\Lambda_{0}\cap\Lambda_{-}} \min\{\mu_{0}(\lambda),\mu_{-}(\lambda) \} d\lambda \right]^2 \right. \nonumber \\ 
    \left. + \left[\int_{\Lambda_{1}\cap\Lambda_{+}} \min\{\mu_{1}(\lambda),\mu_{+}(\lambda) \} d\lambda \right]^2 +\left[\int_{\Lambda_{1}\cap\Lambda_{-}} \min\{\mu_{1}(\lambda),\mu_{-}(\lambda) \} d\lambda \right]^2
    \right\}.
\end{align}

For simpler notation, define 
\begin{align}
    \beta_{0+}\coloneqq\int_{\Lambda_{0}\cap\Lambda_{+}} \min\{\mu_{0}(\lambda),\mu_{+}(\lambda)\} d\lambda, \quad
    \beta_{0-}\coloneqq\int_{\Lambda_{0}\cap\Lambda_{-}} \min\{\mu_{0}(\lambda),\mu_{-}(\lambda)\} d\lambda, \\ 
    \beta_{1+}\coloneqq\int_{\Lambda_{1}\cap\Lambda_{+}} \min\{\mu_{1}(\lambda),\mu_{+}(\lambda)\} d\lambda, \quad
    \beta_{1-}\coloneqq\int_{\Lambda_{1}\cap\Lambda_{-}} \min\{\mu_{1}(\lambda),\mu_{-}(\lambda)\} d\lambda. 
\end{align}
The expression of CE is thus simplified to:

\begin{align}
\label{eq:simoptimal}
    {\rm CE} \leq 1 - \frac{1}{16} \left[
    \beta_{0+}^2 
    + \beta_{0-}^2 + \beta_{1+}^2 +\beta_{1-}^2
    \right].
\end{align}
From Cauchy--Schwarz inequality, we know that 
\begin{align}
    \big( \beta_{0+}^2+\beta_{0-}^2+\beta_{1+}^2+\beta_{1-}^2\big) \big( 1^2+1^2+1^2+1^2\big) \geq  \big(\beta_{0+}+\beta_{0-}+\beta_{1+}+\beta_{1-}\big)^2.
\end{align}
Thus, 
\begin{align} \label{eq:sumbeta}
    {\rm CE} \leq 1 - \frac{1}{64} \big(\beta_{0+}+\beta_{0-}+\beta_{1+}+\beta_{1-}\big)^2.
\end{align}

Let us consider $\beta_{0+}$ first. In general, for a $\lambda\in\Lambda_{0}\cap\Lambda_{+}$, it may also belong to $\Lambda_{1}$ and (or) $\Lambda_{-}$. As such, the ontic space of $\Lambda_{0}\cap\Lambda_{+}$ can be separated into the following four disjoint subregions:
\begin{itemize}
    \item the region where all four ontic spaces overlap, denoted by $\Lambda_{01+-}\coloneqq\Lambda_{0}\cap\Lambda_{1}\cap\Lambda_{+}\cap\Lambda_{-}$;
    \item the region where only $\Lambda_{0}$, $\Lambda_{+}$ and $\Lambda_{1}$ overlap, denoted by $\Lambda_{01+}\coloneqq\Lambda_{0}\cap\Lambda_{1}\cap\Lambda_{+}\setminus\Lambda_{01+-}$;
    \item the region where only $\Lambda_{0}$, $\Lambda_{+}$ and $\Lambda_{-}$ overlap, denoted by $\Lambda_{0+-}\coloneqq\Lambda_{0}\cap\Lambda_{+}\cap\Lambda_{-}\setminus\Lambda_{01+-}$;
    \item the region where  only $\Lambda_{0}$ and $\Lambda_{+}$ overlap, denoted by $\Lambda_{0+}\coloneqq \Lambda_{0}\cap\Lambda_{+} \setminus (\Lambda_{01+-} \cup \Lambda_{01+} \cup \Lambda_{01-}) $. 
\end{itemize}
Since $\Lambda_{0}\cap\Lambda_{+}=\Lambda_{0+}\cup\Lambda_{0+-}\cup\Lambda_{01+}\cup\Lambda_{01+-}$ and since the four subregions are disjoint, 
$\beta_{0+}$ can be expanded as:
\begin{align}
    \beta_{0+} 
    =&\int_{\Lambda_{0+}} \min\{\mu_{0},\mu_{+} \} d\lambda 
    + \int_{\Lambda_{0+-}} \min\{\mu_{0},\mu_{+} \} d\lambda
    + \int_{\Lambda_{01+}} \min\{\mu_{0},\mu_{+} \} d\lambda
    + \int_{\Lambda_{01+-}} \min\{\mu_{0},\mu_{+} \} d\lambda.
\end{align}
Here and later in the proof, the arguments of the $\mu$s are omitted in integrals for concise expressions, i.e. the expression $\mu_{i}$ should be understood as $\mu_{i}(\lambda)$ with $i\in\{0,1,+,-\}$.

Similarly, we can  define 
\begin{align}
    \Lambda_{01-}\coloneqq\Lambda_{0}\cap\Lambda_{1}\cap\Lambda_{-}\setminus\Lambda_{01+-}, \\ \Lambda_{1+-}\coloneqq\Lambda_{1}\cap\Lambda_{+}\cap\Lambda_{-}\setminus\Lambda_{01+-},\\
    \Lambda_{0-}\coloneqq \Lambda_{0}\cap\Lambda_{-} \setminus (\Lambda_{01+-} \cup \Lambda_{01-} \cup \Lambda_{0+-}), \\
    \Lambda_{1+}\coloneqq \Lambda_{1}\cap\Lambda_{+} \setminus (\Lambda_{01+-} \cup \Lambda_{01+} \cup \Lambda_{1+-}), \\ \Lambda_{1-}\coloneqq \Lambda_{1}\cap\Lambda_{-} \setminus (\Lambda_{01+-} \cup \Lambda_{01-} \cup \Lambda_{1+-}), 
\end{align}
and expand the integral regions in $\beta_{0-}$, $\beta_{1+}$ and $\beta_{1-}$ to get:
\begin{align}
    & \beta_{0-} = \int_{\Lambda_{0-}} \min\{\mu_{0},\mu_{-} \} d\lambda 
    + \int_{\Lambda_{0+-}} \min\{\mu_{0},\mu_{-} \} d\lambda
    + \int_{\Lambda_{01-}} \min\{\mu_{0},\mu_{-} \} d\lambda
    + \int_{\Lambda_{01+-}} \min\{\mu_{0},\mu_{-} \} d\lambda, 
    \\ 
    &\beta_{1+} = \int_{\Lambda_{1+}} \min\{\mu_{1},\mu_{+}\} d\lambda 
    + \int_{\Lambda_{1+-}} \min\{\mu_{1},\mu_{+} \} d\lambda
    + \int_{\Lambda_{01+}} \min\{\mu_{1},\mu_{+} \} d\lambda
    + \int_{\Lambda_{01+-}} \min\{\mu_{1},\mu_{+} \} d\lambda, 
    \\ 
    & \beta_{1-} = \int_{\Lambda_{1-}} \min\{\mu_{1},\mu_{-}\} d\lambda 
    + \int_{\Lambda_{1+-}} \min\{\mu_{1},\mu_{-} \} d\lambda
    + \int_{\Lambda_{01-}} \min\{\mu_{1},\mu_{-} \} d\lambda
    + \int_{\Lambda_{01+-}} \min\{\mu_{1},\mu_{-} \} d\lambda.
\end{align}

Let us begin by focusing on the first term in each $\beta_T$. Taking the term $\int_{\Lambda_{0+}} \min\{\mu_{0}(\lambda),\mu_{+}(\lambda) \} d\lambda$ in $\beta_{0+}$ as an example. For any $\lambda\in\Lambda_{0+}$, we have $\mu_{1}(\lambda)=\mu_{-}(\lambda)=0$. Together with \cref{onid}, this implies that $\mu_{0}(\lambda)=\mu_{+}(\lambda)$. Thus $\forall\lambda\in\Lambda_{0+}$, we have $\min\{\mu_{0}(\lambda),\mu_{+}(\lambda) \}=\mu_{0}(\lambda)=\mu_{+}(\lambda)$. Hence,
\begin{align}
\int_{\Lambda_{0+}} \min\{\mu_{0}(\lambda),\mu_{+}(\lambda) \} d\lambda = \int_{\Lambda_{0+}} \mu_{0}(\lambda) d\lambda.
\end{align}

Similarly, \cref{onid} also implies that,
\begin{align}
    \forall \lambda\in\Lambda_{0-},\, \mu_{0}(\lambda)=\mu_{-}(\lambda); \quad 
    \forall \lambda\in\Lambda_{1+},\, \mu_{1}(\lambda)=\mu_{+}(\lambda);\quad
    \forall \lambda\in\Lambda_{1-},\, \mu_{1}(\lambda)=\mu_{-}(\lambda). 
\end{align}
Thus, the expressions of the $\beta_T$s are simplified to:
\begin{align}
    &\beta_{0+} = \int_{\Lambda_{0+}}\mu_{0}d\lambda 
    + \int_{\Lambda_{0+-}} \min\{\mu_{0},\mu_{+} \} d\lambda + \int_{\Lambda_{01+}} \min\{\mu_{0},\mu_{+} \} d\lambda + \int_{\Lambda_{01+-}} \min\{\mu_{0},\mu_{+} \} d\lambda, \\ 
    & \beta_{0-} = \int_{\Lambda_{0-}} \mu_{0} d\lambda 
    + \int_{\Lambda_{0+-}} \min\{\mu_{0},\mu_{-} \} d\lambda
    + \int_{\Lambda_{01-}} \min\{\mu_{0},\mu_{-} \} d\lambda
    + \int_{\Lambda_{01+-}} \min\{\mu_{0},\mu_{-} \} d\lambda,
    \\ 
    &\beta_{1+} = \int_{\Lambda_{1+}}\mu_{1} d\lambda 
    + \int_{\Lambda_{1+-}} \min\{\mu_{1},\mu_{+} \} d\lambda
    + \int_{\Lambda_{01+}} \min\{\mu_{1},\mu_{+} \} d\lambda
    + \int_{\Lambda_{01+-}} \min\{\mu_{1},\mu_{+} \} d\lambda, 
    \\ 
    & \beta_{1-} = \int_{\Lambda_{1-}} \mu_{1} d\lambda 
    + \int_{\Lambda_{1+-}} \min\{\mu_{1},\mu_{-} \} d\lambda
    + \int_{\Lambda_{01-}} \min\{\mu_{1},\mu_{-} \} d\lambda
    + \int_{\Lambda_{01+-}} \min\{\mu_{1},\mu_{-} \} d\lambda.
\end{align}

For the second and the third terms in the $\beta_T$s, the integral regions are the ones overlapped by three out of the four ontic spaces in $\{\Lambda_0,\Lambda_1,\Lambda_+,\Lambda_-\}$. There are eight terms in total, but only four different integral regions. Let us first look at the terms that integrate over $\Lambda_{0+-}$, which are the second terms in $\beta_{0+}$ and $\beta_{0-}$. From \cref{onid}, for any $\lambda\in\Lambda_{0+-}$, we have $\mu_{0}(\lambda)=\mu_{+}(\lambda)+\mu_{-}(\lambda)$, and thus, $\mu_{0}(\lambda) \geq \mu_{+}(\lambda)$ and $\mu_{0}(\lambda) \geq \mu_{-}(\lambda)$. Therefore, the sum of the second terms of $\beta_{0+}$ and $\beta_{0-}$ (recall from \cref{eq:sumbeta} that we need to sum all $\beta_T$s) satisfies
\begin{align}
    \int_{\Lambda_{0+-}}[\min\{\mu_{0}(\lambda),\mu_{+}(\lambda)\}+\min\{\mu_{0}(\lambda),\mu_{-}(\lambda)\}] d\lambda
    =\int_{\Lambda_{0+-}}[\mu_+(\lambda)+\mu_-(\lambda)]d\lambda
    =\int_{\Lambda_{0+-}}[\mu_0(\lambda)+\mu_1(\lambda)]d\lambda,
\end{align}
where in the last line we used \cref{onid} again. 

Similarly, for the other six terms where the integral regions are overlapped by three out of the four ontic states, we can derive that,
\begin{align}
\label{eq:3overlap}
    \int_{\Lambda_{1+-}} [\min\{\mu_{1},\mu_{+}\}+\min\{\mu_{1},\mu_{-}\}] d\lambda
    =&\int_{\Lambda_{1+-}} [\mu_0+\mu_1] d\lambda,\\ \nonumber
    \int_{\Lambda_{01+}} [\min\{\mu_{0},\mu_{+}\}+\min\{\mu_{1},\mu_{+}\}] d\lambda
    =&\int_{\Lambda_{01+}} [\mu_0+\mu_1] d\lambda,\\ \nonumber
    \int_{\Lambda_{01-}} [\min\{\mu_{0},\mu_{-}\}+\min\{\mu_{1},\mu_{-}\}] d\lambda
    =&\int_{\Lambda_{01-}} [\mu_0+\mu_1] d\lambda.\\ \nonumber
\end{align}
Therefore, when we sum all the $\beta_T$s together, we have:
\begin{align}
\label{eq:betamid}
    &\beta_{0+}+\beta_{0-}+\beta_{1+}+\beta_{1-} \\ \nonumber
    =& \int_{\Lambda_{0+}} \mu_{0} d\lambda + \int_{\Lambda_{0-}} \mu_{0} d\lambda + \int_{\Lambda_{1+}} \mu_{1} d\lambda + \int_{{\Lambda_{1-}}} \mu_{1} d\lambda \\ \nonumber
    &+ \int_{\Lambda_{0+-}} [\mu_0+\mu_1] d\lambda + \int_{\Lambda_{1+-}} [\mu_0+\mu_1] d\lambda +\int_{\Lambda_{01+}} [\mu_0+\mu_1] d\lambda +\int_{\Lambda_{01-}} [\mu_0+\mu_1] d\lambda \\ \nonumber
    & + \int_{\Lambda_{01+-}} 
    [\min\{\mu_{0},\mu_{+} \} + \min\{\mu_{0},\mu_{-} \} + \min\{\mu_{1},\mu_{+} \} + \min\{\mu_{1},\mu_{-}\} ] d\lambda, \\ \nonumber
\end{align}
where the last term is obtained by combining the forth terms from each $\beta_T$.

The integral region in the last term in Eq.~\eqref{eq:betamid} is overlapped by all four ontic spaces. If, for example, for some $\lambda\in\Lambda_{01+-}$, $\mu_{+}(\lambda)$ is the smallest among $\{\mu_{0}(\lambda),\mu_{1}(\lambda),\mu_{+}(\lambda),\mu_{-}(\lambda) \}$, \cref{onid} then implies that $\mu_{-}(\lambda)$ is the largest among the four. Therefore, for such a $\lambda$, the expression in the integral in the last term satisfies
\begin{align}
 & \min\{\mu_{0}(\lambda),\mu_{+}(\lambda) \} + \min\{\mu_{0}(\lambda),\mu_{-}(\lambda) \} + \min\{\mu_{1}(\lambda),\mu_{+}(\lambda) \} + \min\{\mu_{1}(\lambda),\mu_{-}(\lambda) \} \\ \nonumber
 =&
 \mu_{+}(\lambda)+\mu_{0}(\lambda) + \mu_{+}(\lambda)+\mu_{1}(\lambda). \\ \nonumber
\end{align}
By repeating the analyses for the case where $\mu_{-}$, $\mu_{0}$ or $\mu_{1}$ is the smallest among the four, we see that 
\begin{align}
\nonumber
 & \min\{\mu_{0}(\lambda),\mu_{+}(\lambda) \} + \min\{\mu_{0}(\lambda),\mu_{-}(\lambda) \} + \min\{\mu_{1}(\lambda),\mu_{+}(\lambda) \} + \min\{\mu_{1}(\lambda),\mu_{-}(\lambda) \} \\ \nonumber
 =& \mu_0(\lambda)+\mu_1(\lambda) + 2\min \{\mu_{0}(\lambda),\mu_{1}(\lambda),\mu_{+}(\lambda),\mu_{-}(\lambda) \} 
\end{align}

Therefore, the expression of the sum of the $\beta_T$s can be further simplified to
\begin{align}
    &\beta_{0+}+\beta_{0-}+\beta_{1+}+\beta_{1-} \\ \nonumber
    =& \int_{\Lambda_{0+}} \mu_{0} d\lambda + \int_{\Lambda_{0-}} \mu_{0} d\lambda + \int_{\Lambda_{1+}} \mu_{1} d\lambda + \int_{{\Lambda_{1-}}} \mu_{1} d\lambda \\ \nonumber
    &+ \int_{\Lambda_{0+-}} [\mu_0+\mu_1] d\lambda + \int_{\Lambda_{1+-}} [\mu_0+\mu_1] d\lambda +\int_{\Lambda_{01+}} [\mu_0+\mu_1] d\lambda +\int_{\Lambda_{01-}} [\mu_0+\mu_1] d\lambda \\ \nonumber
    & + \int_{\Lambda_{01+-}} 
    [\mu_0+\mu_1 + 2\min \{\mu_{0},\mu_{1},\mu_{+},\mu_{-}\}] d\lambda. \\ \nonumber
\end{align}
Since $\min \{\mu_{0},\mu_{1},\mu_{+},\mu_{-}\}\geq0$, we can remove the last term in the last integral and get the inequality:
\begin{align}
    &\beta_{0+}+\beta_{0-}+\beta_{1+}+\beta_{1-} \\ \nonumber    
    \geq & \int_{\Lambda_{0+}} \mu_{0} d\lambda + \int_{\Lambda_{0-}} \mu_{0} d\lambda + \int_{\Lambda_{1+}} \mu_{1} d\lambda + \int_{{\Lambda_{1-}}} \mu_{1} d\lambda \\ \nonumber
    &+ \int_{\Lambda_{0+-}} [\mu_0+\mu_1] d\lambda + \int_{\Lambda_{1+-}} [\mu_0+\mu_1] d\lambda +\int_{\Lambda_{01+}} [\mu_0+\mu_1] d\lambda +\int_{\Lambda_{01-}} [\mu_0+\mu_1] d\lambda \\ \nonumber
    & + \int_{\Lambda_{01+-}} 
    [\mu_0+\mu_1 ] d\lambda, \\ \nonumber
\end{align}
By regrouping the terms in the above expression, we get
\begin{align}
    &\beta_{0+}+\beta_{0-}+\beta_{1+}+\beta_{1-} \\ \nonumber
    \geq &\int_{\Lambda_{0+}\cup\Lambda_{0-}\cup\Lambda_{0+-}\cup\Lambda_{01+}\cup\Lambda_{01-}\cup\Lambda_{01+-}} \mu_{0} d\lambda +  \int_{\Lambda_{1+}\cup\Lambda_{1-}\cup\Lambda_{1+-}\cup\Lambda_{01+}\cup\Lambda_{01-}\cup\Lambda_{01+-}} \mu_{1} d\lambda\\ \nonumber
    =&\int_{\Lambda_{0}} \mu_{0} d\lambda +  \int_{\Lambda_{1}} \mu_{1} d\lambda\\ \nonumber
    =&2
\end{align}

As such, from \cref{eq:sumbeta}, we have that 
\begin{align}
    {\rm CE}\leq_{\rm nc} 1-\frac{1}{64}2^2=93.75\%.
\end{align}

Now we show that this bound is tight by constructing a quantum experiment saturating the bound while admitting of a noncontextual ontological model.
Consider the experiment where the sets of states $\mathcal{P}_{T}$ are the same as the quantum ones in the main text, e.g., $\mathcal{P}_{0+}=\{\ket{00}, \ket{0+}, \ket{+0}, \ket{++}\}$, while the measurements are different from the ones in the main text. Specifically, now $\mathcal{M}_{0+}=\mathcal{M}_{0-}= \{ \ket{11}, \ket{10}, \ket{01}, \ket{00}\}$,  and $\mathcal{M}_{1+}= \mathcal{M}_{1-}=\{ \ket{00}, \ket{01}, \ket{10}, \ket{11}\}$. Then, the conclusiveness of exclusion in each of the four tasks is
\begin{align}
	{\rm CE}_{0+}={\rm CE}_{0-}={\rm CE}_{1+}={\rm CE}_{1-}= 1- \frac{1}{4}(0 + 0+ 0+ \frac{1}{4}) =\frac{15}{16}.
\end{align}
Thus, 
\begin{align}
	{\rm CE}=93.75\%.
\end{align}
This experiment admits a noncontextual ontological model because all states and measurements considered here have exact analogues in Spekkens’ toy theory~\cite{spekkens2007evidence}, which admits a noncontextual ontological model.

\end{widetext}

\section{From the PBR construction to the bilocality scenario}\label{app:bilocality}

One natural way to implement each of the multi-sources in \cref{fig:table} is via quantum steering, as depicted in Fig.~\ref{BilocalityScenario}. 
This is possible because of the operational identity of Eq.~\eqref{opeq}, $\frac{1}{2} |0\rangle\langle 0| +\frac{1}{2} |1\rangle\langle 1| = \frac{1}{2} |+\rangle\langle +| +\frac{1}{2} |-\rangle\langle -|$ which expresses the fact that the $Z$-source and $X$-source produce the same average quantum state if one marginalizes over their outcome, i.e., the condition of no-signaling from $S_A$ to $A$ and from $S_B$ to $B$, which must be satisfied for the causal structure assumed in Fig.~\ref{BilocalityScenario}. This causal structure is known as the bilocality scenario~\cite{bilocality}.

The bilocality scenario induces a prepare-measure scenario on the bipartite system $AB$ via steering, and so we can relate it to our conclusive exclusion task from the main text. This is analogous to a well-known method of relating noncontextuality no-go theorems in prepare-measure experiments with locality no-go theorems in Bell scenarios~\cite{Speckersparable,schmidContextual2018,schmidAll2018,Wright2023invertible}. The connection is straightforward.  The effective device obtained by composing the entangled state $A'A$ with the multi-meter on $A'$ is a multi-source on $A$.  Similarly for $B$.  This composition is illustrated by the red-dashed boxes in Fig.~\ref{BilocalityScenario}. The operational identity which was assumed to hold for the multi-source in our conclusive exclusion task in, i.e., \cref{opeq}, is now a necessary consequence of the causal structure, as mentioned above. Moreover, the ontological identity that follows from this condition together with the assumption of noncontextuality, namely $\frac{1}{2} \mu_0(\lambda) +\frac{1}{2} \mu_1(\lambda) = \frac{1}{2} \mu_+(\lambda) +\frac{1}{2} \mu_-(\lambda)$, is now a constraint on any classical causal model with this causal structure. 

By the same logic as in the main text, then, it follows that there is no classical causal model within the bilocality scenario that can reproduce the perfect conclusive exclusion that quantum theory allows. 
It also follows that the inequality in Eq.~\eqref{NCinequality} is a causal compatibility inequality in the bilocality scenario---that is, an inequality that follows for any classical causal model having the causal structure depicted in Fig.~\ref{fig:setup}.
Finally, it follows that one can find a quantum violation of this inequality in the bilocality scenario.  

Among proofs of quantum-classical gaps in the bilocality scenario that do not piggyback on Bell's theorem by leveraging entanglement swapping, this is to our knowledge the first example of a possibilistic such proof.

\end{document}